\documentclass[useAMS,usenatbib,twocolumn]{mn2e}
\usepackage{amsfonts}
\usepackage{amsmath}
\usepackage[dvips]{graphicx}
\usepackage{epsfig}
\usepackage{enumitem}   
\usepackage{wrapfig}
\usepackage{amssymb}
\usepackage{longtable}
\usepackage{accents}
\usepackage{dcolumn}
\usepackage{multirow}
\usepackage{bm}
\usepackage{url}
\usepackage[caption=false]{subfig}

\begin{document}

\title[Candidate for a track-type neutrino event]{A flat spectrum candidate for a track-type high energy neutrino emission event, the case of blazar PKS~0723--008}
\author[E. Kun, P. L. Biermann, L. \'{A}.
Gergely]{E. Kun$^{1}$\thanks{%
E-mail: kun@titan.physx.u-szeged.hu}, P. L. Biermann$^{2,3,4,5}$, L. \'{A}. Gergely$^{1}$ \\
$^{1}$ Institute of Physics, University of Szeged, D\'om t\'er 9, H-6720 Szeged, Hungary\\
$^{2}$ Max Planck Institute for Radioastronomy, Auf dem H\"{u}gel 69, D-53121 Bonn, Germany\\
$^{3}$ Department of Physics, Karlsruhe Institute for Technology, P.O. Box 3640, D-76021, Karlsruhe, Germany\\
$^{4}$ Department of Physics \& Astronomy, University of Alabama, AL 35487-0324, Tuscaloosa, USA\\
$^{5}$ Department of Physics \& Astronomy, University of Bonn, Regina-Pacis-Weg 3, 53113, Bonn, Germany}
\date{Accepted . Received ; in original form }
\maketitle

\begin{abstract}
By cross-correlating both the Parkes Catalogue and the Second Planck Catalogue of Compact Sources with the arrival direction of the track-type neutrinos detected by the IceCube Neutrino Observatory, we find the flat-spectrum blazar PKS 0723--008 as a good candidate for the high-energy neutrino event 5 (ID5). Apart from its coordinates matching those of ID5, PKS~0723--008 exhibits further interesting radio properties. Its spectrum is flat up to high \textit{Planck} frequencies, and it produced a fivefold-increased radio flux density through the last decade. Based upon these radio properties we propose a scenario of binary black hole evolution leading to the observed high-energy neutrino emission. The main contributing events are the spin-flip of the dominant black hole, the formation of a new jet with significant particle acceleration and interaction with the surrounding material, with the corresponding increased radio flux. Doppler boosting from the underlying jet pointing to the Earth makes it possible to identify the origin of the neutrinos, so the merger itself is the form of an extended flat-spectrum radio emission, a key selection criterion to find traces of this complex process.
\end{abstract}

\pagerange{\pageref{firstpage}--\pageref{lastpage}}

\label{firstpage}


\section{Introduction}

Up to date there is no compelling understanding on the origin of extraterrestrial high-energy (HE) neutrinos observed by the IceCube Neutrino Observatory \citep{IC2014,IC2015}. Recently \citet{Kadler2016} reported that a major outburst of the blazar PKS B1424--418 occurred in temporal and positional coincidence with the third PeV-energy neutrino event (IC35) detected by the IceCube Neutrino Observatory. They have shown, that the outburst of PKS B1424--418 provided an energy output high enough to explain the observed PeV event, indicative of a direct physical association.

In this Letter we present a promising candidate for another HE-neutrino event (ID5), the flat-spectrum blazar PKS~0723--008. In contrast to PKS~B1424--418, which is a shower-type event, this source is candidate for a track-type event. Shower-type events exhibit spherical topology, and they are created by the electrons emerging from the interaction of the electron-neutrinos with the ice. Such electrons scatter several times until their energy fall below the Cherenkov threshold. By contrast, track-type neutrino events appear as longer tracks, generated by the muons emerging from the interaction of muon-neutrinos with ice. For this type of HE neutrino events the average median angular error is $\lesssim 1.\degr2$, being about 10 times smaller than for the shower-type event ID35 ($15.\degr9$).

We also propose a scenario explaining both the HE neutrino emission and the radio properties of PKS~0723--008, based on the merger of two supermassive black holes (SMBH). Typical SMBH evolution leads to a spin-flip of the dominant relativistic jet \citep{Gergely2009}, followed by emission of low-frequency gravitational waves. After the spin-flip a new channel must be plowed through the surrounding material \citep{Becker2009}. At the end of this process, as the jet penetrates the outer regions of the host galaxy, it will accelerate ultrahigh-energy cosmic ray (UHECR) particles, usually protons, therefore substantial UHECR emission is expected after the merger. The nature and origin of UHECR particles were recently reviewed in \citet{Biermann2016}. The UHECR-background has been detected by the Pierre Auger Observatory \citep{Auger2009}, KASCADE-Grande \citep{KG2011} and Telescope Array experiment \citep{TA2014} at EeV energies. Protons get accelerated to very HE in radio galaxies and their jets at about $10^{20}$ eV, based on observed active galactic nucleus (AGN) spectra \citep[e.g][]{Aharonian2002}. A simple estimate leads to $10^{21}$ eV as the maximum in the jet frame \citep{Biermann1987}. Then, if the proton energy is above pion-production threshold, proton-proton hadronic collisions produce pions that decay further creating neutrinos with PeV energies \citep{Kadler2016}.


Optical spectroscopic and very long baseline interferometry observations on the inner part of some AGN are consistent with a presence of a spatially unresolved SMBH binaries. Such sources are e.g. Mrk~501 \citep{Villata1999}, 3C~273 \citep{Romero2000}, BL~Lac \citep{Stirling2003}, 3C~120 \citep{Caproni2004}, S5~1803+784 \citep{Roland2008}, NGC~4151 \citep{Bon2012}, S5~1928+738 \citep{Roos1993,Kun2014}, PG~1302--102 \citep{Graham2015,Kun2015}. The observation-compatible binary parameters typically imply the inspiral phase of the binary evolution. Such systems may be revealed in near-THz radio or follow-up HE neutrino emission observations, as will be discussed in detail.

This Letter is organized as follows. In Section \ref{section2}, we describe the process of the candidate selection based on the Parkes Catalogue and the Second Planck Catalogue of Compact Sources. In Section \ref{section3} we present the most promising candidate of a source responsible for the HE neutrino. In Section \ref{section4} we give more details on the scenario leading to the HE neutrino emission, while in Section \ref{section5} we briefly discuss and summarize our results.

\section{On the origin of the extraterrestrial neutrinos -- candidate selection}
\label{section2}

From the $54$ extragalactic neutrino detections by the IceCube Neutrino Observatory $14$ of them are track-type events with median angular error of $1.\degr2$: IDs $3$, $5$, $8$, $13$, $18$, $23$, $28$, $37$, $38$, $43$, $44$, $45$, $47$, $53$ \citep{IC2014,IC2015}, and a 15th ID with a total energy of $2.9\pm0.3$PeV \citep{Schoenen2015}. The question is how to account these neutrinos to their origin.

\begin{table*}
\caption{Details of the source detection from the Second \textit{Planck} compact source catalogue. (1) Source name in the PCCS2 catalogue, (2) frequency, (3) J$2000$ right-ascension, (4) J$2000$ declination, (5) Galactic latitude, (6) Galactic longitude, (7) flux density as determined by aperture photometry, (8) detection flux, (9) flux density as determined by PSF photometry, (10) flux density as determined by Gaussian photometry. Due to its closeness and brightness we identify this source as being the flat-spectrum blazar PKS~0723-008.}
\label{table:planck_fluxes}
\resizebox{\textwidth}{!}{\begin{tabular}{cccccccccc}
\hline
Source ID & $\nu$ & RA & Dec. & $b$ & $l$ & $S_\mathrm{aper}$ & $S_\mathrm{det}$ & $S_\mathrm{PSF}$ & $S_\mathrm{Gauss}$ \\
 & (GHz) & ($\degr$) & ($\degr$) & ($\degr$) & ($\degr$) & (mJy) & (mJy) & (mJy) & (mJy) \\
(1) & (2) & (3) & (4) & (5) & (6) & (7) & (8) & (9) & (10) \\
\hline
PCCS2 030 G217.71+07.23	& $30$  & $111.47$ & $-0.92$ & $7.23$ & $217.71$ & $3839\pm476$	& $4341\pm 96$	& $3765 \pm 196$	& $3835	\pm 47$ \\
PCCS2 044 G217.72+07.23	& $44$ & $111.47$ & $-0.93$ & $7.23$ & $217.71$ & $3845\pm	758$ &$4238\pm 165$	& $3654	\pm 362$	& $3581	\pm 27$\\
PCCS2 070 G217.70+07.23	& $70$ & $111.46$ & $-0.92$ & $7.23$ & $217.70$ & $4080\pm	298$ &$4009\pm 110$	& $3899	\pm 593$	& $3877	\pm 38$\\
PCCS2 100 G217.68+07.21	& $100$ & $111.45$ & $-0.91$ & $7.22$ & $217.68$ & $3106\pm	227$ &$3174\pm 61$	& $3138	\pm 231$	& $3195	\pm 47$\\
PCCS2 143 G217.68+07.22	& $143$ & $111.46$ & $-0.91$ & $7.23$ & $217.69$ & $2640\pm	157$ &$2662\pm 44$	& $2700	\pm 323$	& $2722	\pm 47$\\
PCCS2 217 G217.70+07.23	& $217$ & $111.47$ & $-0.92$ & $7.24$ & $217.70$ & $2197\pm	109$ &$2374\pm 41$	& $2334	\pm 439$	& $2171	\pm 64$\\
PCCS2E 353 G217.71+07.22 & $353$ & $111.47$ & $-0.93$ & $7.22$ & $217.71$ & $1758\pm	252$ &$2056\pm 73$	& $1743 \pm 538$	& $1699 \pm 157$\\
PCCS2E 545 G217.71+07.20 & $545$ & $111.45$ & $-0.94$ & $7.21$ & $217.71$ & $1388\pm	569$ &$1799\pm 112$	& $1141 \pm 1132$	& $1526 \pm 81$\\
PCCS2E 857 G217.68+07.21 & $857$ & $111.44$ & $-0.91$ & $7.21$ & $217.69$ & $1675\pm	839$ &$1437\pm 206$	& $348	\pm 2987$	& $2092 \pm 282$\\
\hline
\end{tabular}}
\end{table*}
 
  \begin{figure*}
 \includegraphics[scale=0.38]{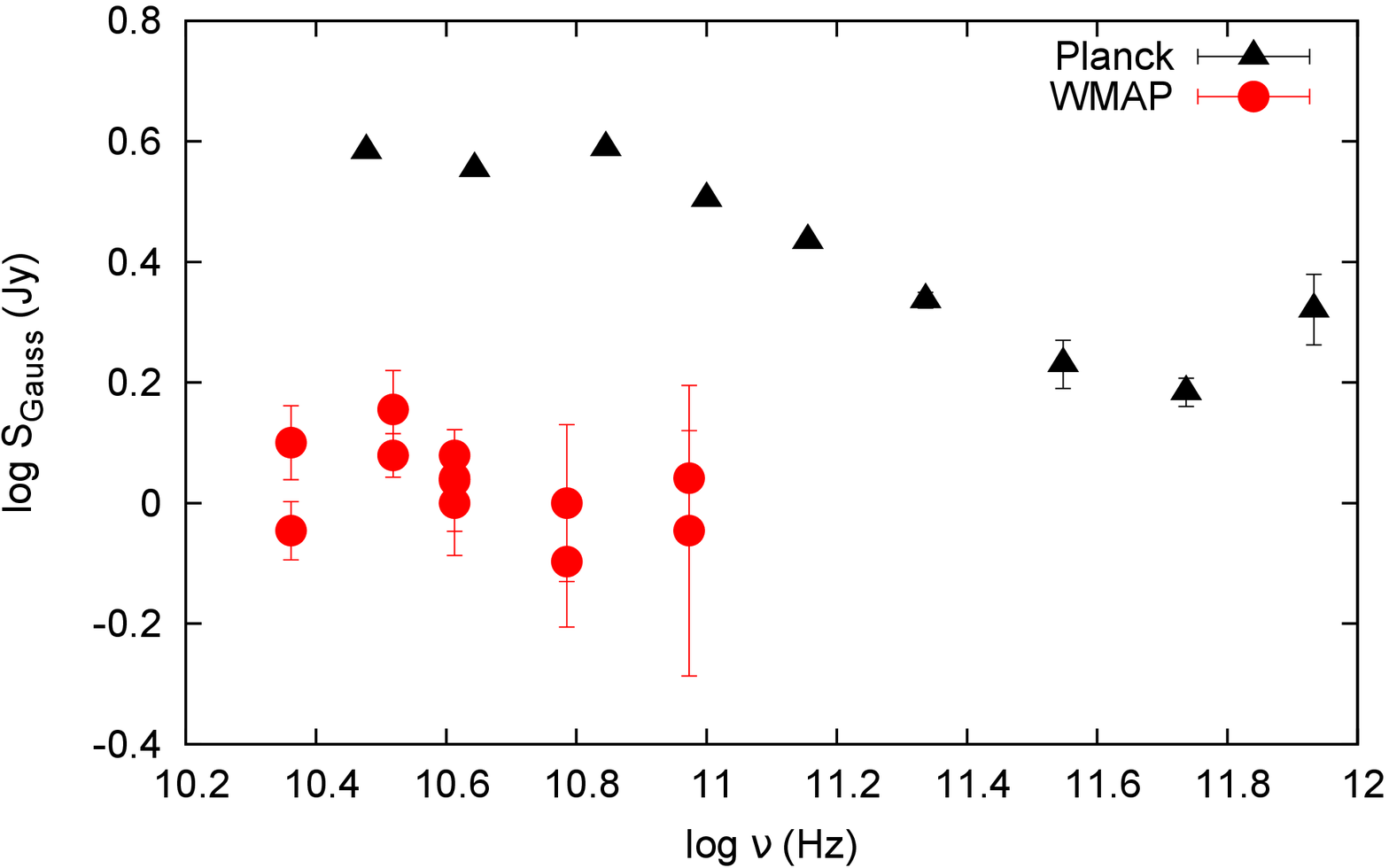}
  \includegraphics[scale=0.38]{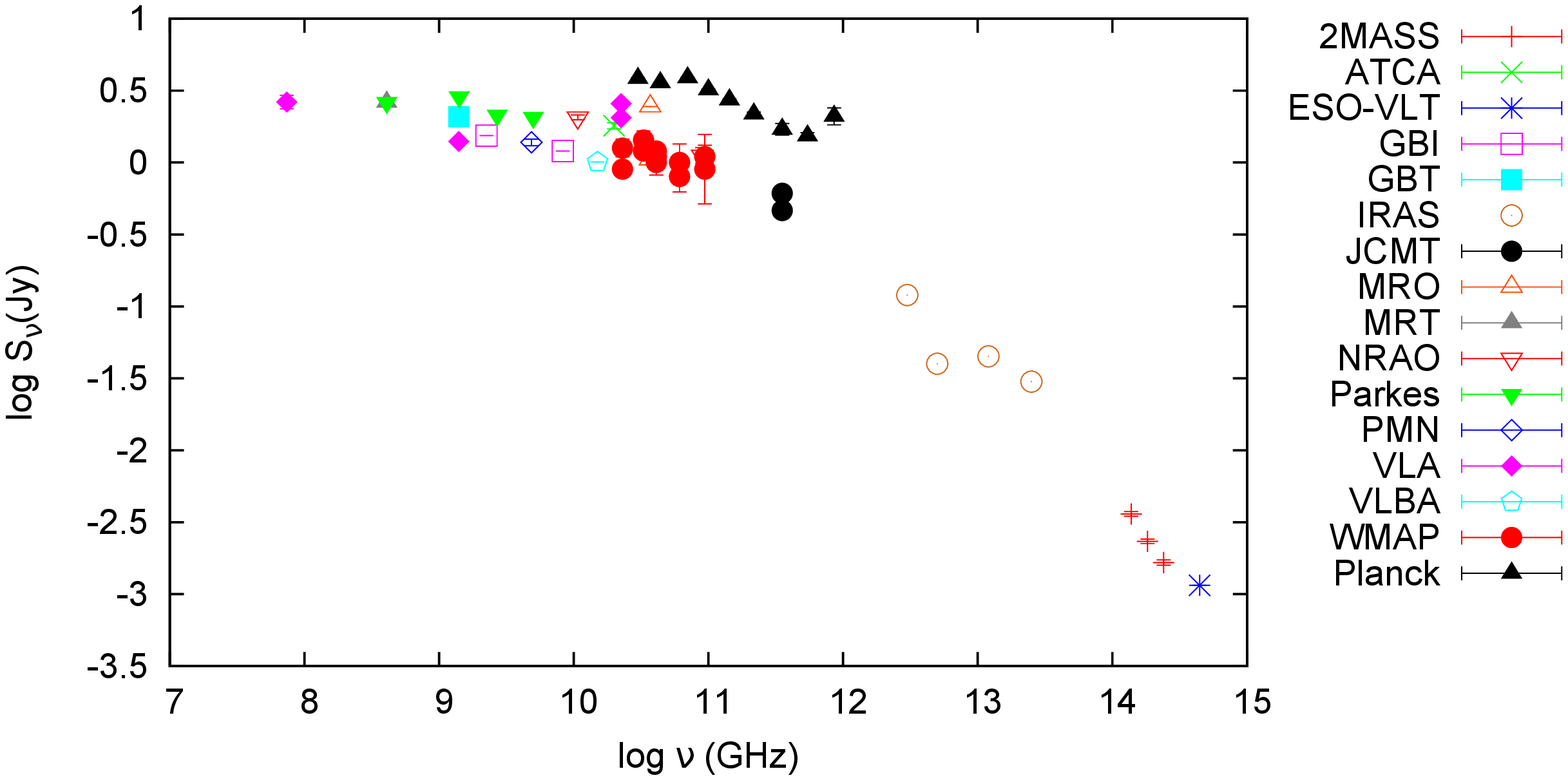}
 \caption{The left-hand panel shows the \textit{Planck} (listed in Table \ref{table:planck_fluxes}) and \textit{WMAP} spectrum of the candidate source. The spectral index is $\alpha_\mathrm{30GHz, 857GHz}=-0.18\pm0.04$, consistent with the criterion of \citet{Gregorini1984} of assuming $-0.5$ as a dividing line in spectra with the convention $S(\nu)\sim\nu^\alpha$. The steepest part of the \textit{Planck} spectrum is still slightly flat by this criterion, $\alpha_\mathrm{70GHz, 545GHz}=-0.45\pm0.03$. The right-hand panel shows the available spectral informations of PKS~0723--008 taken from the NASA/IPAC Extragalactic Database.}
 \label{figure:planck_spectra}
 \end{figure*}
 
 \begin{figure*}
 \includegraphics[scale=0.3]{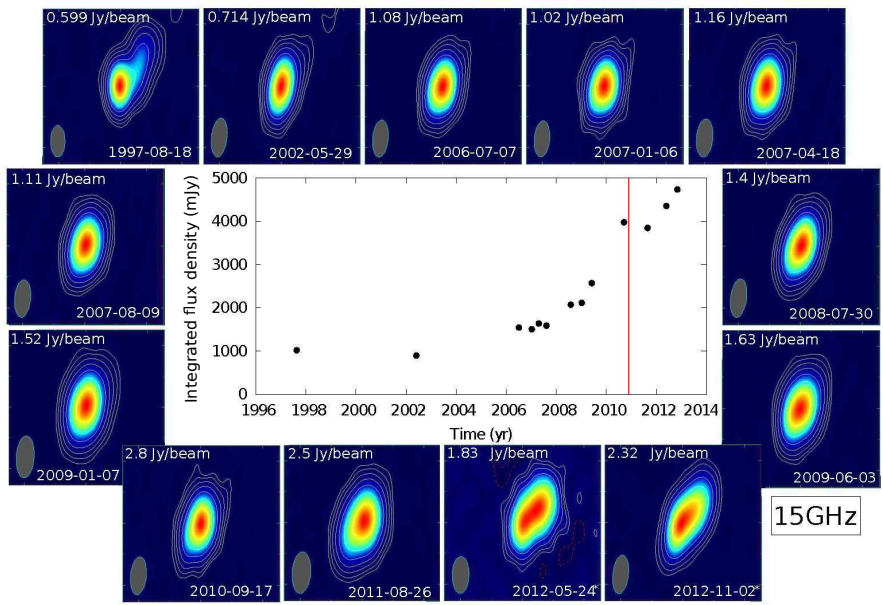}
 \caption{The radio maps of PKS~0723-008 over $12$ epochs, represented on logarithmic scale with base ten. They were produced by processing the available VLBA visibilities provided by the MOJAVE team. Iso-flux density contours are in per cent of the peak flux density marked in the left upper corner of the maps. They increase by factors of $1$, except the last two epochs (marked by stars), where the contours increase by factors of $2$. In the middle the integrated flux density of the source at $15$~GHz is represented as function of the time. The time of the corresponding neutrino detection (ID5) is indicated by a red vertical line.}
 \label{figure:vlba}
 \end{figure*}

 We searched for flat-spectrum sources (flat between $2.7$ and $5$GHz) in the $1$Jy catalogue \citep{Kuhr1981} and in the Parkes Catalogue \citep{PKS1990} within the median angular error of the $14$ track-type neutrino events observed by the IceCube Neutrino Observatory \citep{IC2014,IC2015}. In the 1Jy catalogue we did not find flat-spectrum radio sources in the required proximity. However, in the Parkes catalogue we found four candidates: PKS~0723--008 ($z=0.128$; ID5), PKS~B1206--202 ($z=0.404$; ID8), PKS~B2300--254 ($z$ unknown; ID18), PKS B2224+006=4C+00.81 ($z=2.25$; ID44). Here and thereafter we mark the corresponding HE neutrino events in parentheses.

Then we cross-correlated the Second Planck Catalogue of Compact Sources \citep[hereafter PCCS2;][]{PCCS2015} with the position of the track-type HE neutrino events. Here we found the PCCS2-counterpart of the blazar PKS~0723-008 (ID5) and of the quasar 4C+00.81 (ID44) only. These sources are flat across to the \textit{Planck} frequencies. The spectral index of PKS~0723--008 is $\alpha=-0.29\pm0.02$ between 30 and 217GHz, and $\alpha=-0.18\pm0.04$ between 30 and 857GHz, with the convention $S(\nu)\sim\nu^\alpha$. The spectral index of 4C+00.81 is $\alpha=-0.30\pm0.07$ between 30 and 217GHz, and $\alpha=0.16\pm0.07$ between 30 and 545GHz (we did not find the counterpart of the source at 857GHz). 

The last known flux density of PKS~0723--008 at 15GHz measured by the Monitoring Of Jets in Active galactic nuclei with VLBA Experiments (MOJAVE) team is 4807 mJy \citep{Lister2013}. This is almost 13 times larger than the 15GHz flux density of the 4C+00.81, which is $328$ mJy by the Owens Valley Radio Observatory \citep{Richards2011}, making PKS~0723-008 the best HE neutrino counterpart candidate.

\section{The candidate source: PKS~0723--008}
\label{section3}

The sky-coordinates of the flat-spectrum blazar PKS 0723--008 are RA$_\mathrm{J2000}=+111.\degr4610$ and DEC$_\mathrm{J2000}=-0.\degr9157$, being near the position of HE event ID5. ID5 was observed on 2010 November 12 with energy $\sim 71.4$ TeV and centred about RA$_\mathrm{J2000}=+110.\degr6$, Dec$_\mathrm{J2000}=-0.\degr4$, with median angular error of $\lesssim 1.\degr2$. 

We summarize the corresponding PCCS2 detections in Table \ref{table:planck_fluxes}, and show the spectrum in the left-hand panel of Fig. \ref{figure:planck_spectra}.  
We also show a more detailed spectrum in the right-hand panel of Fig. \ref{figure:planck_spectra}, taken from the NASA/IPAC Extragalactic Database (NED). PKS~0723--008 has flat spectrum towards the high \textit{Planck} frequencies, revealing the most internal component of the jet to be active.

The \textit{Planck} spectrum is brighter than the \textit{Wilkinson Microwave Anisotropy Probe} (\textit{WMAP}) spectrum at the same frequencies. During the \textit{WMAP} measurements (between $2001$ and $2006$) the flux density of the source maintained its integrated flux density about $1000$mJ at $15$GHz, while the Very Large Baseline Array (VLBA) observations reveal that its flux density was increasing after $2006$ (see Fig. \ref{figure:vlba}). 
The \textit{Planck} observed the source after 2006, when it was about four times brighter compared to the constant phase. 

PKS~0723-008 was monitored by the MOJAVE team with the VLBA at $15$GHz across $10$ yr in $12$ epochs \citep{Lister2013}. We cleaned the available calibrated $uv$-visibilities to see if any activity is going on, as the radio spectrum suggests. We decomposed the surface brightness of the source into circular Gaussian components, but no prominent component motion emerged. Standard \textsc{Difmap} tasks \citep{Shepherd1994} were used to clean and model-fit the data.
 
In Fig. \ref{figure:vlba}, we show the integrated flux of the source, along with the cleaned radio maps. Its radio flux has been increasing since $2006$, indicating a strengthening activity. The integrated flux exhibits a local maximum at the end of 2010. On the other hand, a doubled core is visible on the radio map in Fig. \ref{figure:vlba} at epoch 2012-05-24, and a largely elongated core at the next epoch at 2012-11-02. The flux increasing and the restructuring of the core suggest a component ejection.

Fig. \ref{figure:vlba} shows that the radio emission already began to increase before the IceCube detected the ID5 at 2010-11-12, indicating the plowing of a new jet channel already before that.

\section{A possible scenario to explain the HE neutrinos}
\label{section4}

In this section, we present a scenario to produce HE neutrinos. As shown in \citet{Gergely2009} after a merger of two SMBHs, the original orbital angular momentum determines the spin direction of the newly merged SMBH. At the same time the centre of the newly merged host galaxy is full of low angular momentum gas from the mixing of the two galaxies: in the case of originally hot gas, this may undergo a phase transition to cool the temperature in the pressure jumps of the violence of the interstellar medium merging \citep[e.g,][]{Biermann1972,Toomre1972,Biermann1983,Mulchaey2000}. This low angular momentum gas will allow the formation of a newly formed accretion disc, and so maximally feed a powerful jet \citep[e.g,][]{Falcke1995,Donea1996,Araudo2012,Ghisellini2014,Sbarrato2016,Ghisellini2015}.

On the other hand, the gas surrounding the freshly merged central SMBH may be just too agitated, and so too hot for the formation of an accretion disc; in such a case, a jet may be formed directly as in the Blandford--Znajek model \citep{BZ1977}, with the magnetic field assumed to still exist from earlier accretion episodes; such a jet starts as a Poynting flux jet \citep[e.g,][]{Kronberg2011,Tavecchio2016} and so with a very high Lorentz factor; it may acquire its hadronic load by interaction with the surroundings or with stars \citep{Becker2009,Araudo2012,Muller2015,Wykes2015,Romero2016}.

In either case such a jet is then optically thick over a range of frequencies; the maximum frequency over which it is optically thick most strongly depends on the Lorentz factor and boosting of the compact jet \citep[e.g,][]{Blandford1979,Falcke1995}. Focusing on the most relevant dependence in the expression of this maximum frequency we note that the expressions in \citet{Blandford1979} and \citet{Falcke1995} have very nearly the same maximum frequency dependence on the jet Lorentz factor $\gamma_j$ with $\nu_{b} \, \sim \, \gamma_j^2$ at the maximal optimum angle to the line of sight \citep[in][this power is 11/6]{Blandford1979}. Here $\nu_{b}$ is the frequency up to which the spectrum will be flat. This is the upper cut-off frequency of the electron energy distribution evaluated at radius $r_b$, where the maximum brightness temperature is reached. The observed Lorentz factors range from just above unity to nearly 100 \citep[e.g.,][]{Gopal-Krishna2010}. This implies that this maximal frequency is most easily shifted to a high frequency in the case that the Lorentz factor of the jet is high. So we may have the highest Lorentz factors in the second case, with little accretion and a powerful Poynting flux jet initially.

The neutrino emission is due to energetic proton--proton hadronic collisions, where the kinetic energy of the protons is above the energy threshold of pion-creation. The jet, freshly started after the final merger, contains the seed-protons and accelerates them to reach the energy level of the pion production ($p^+ + p^+ \rightarrow p^+ + n^0 +\pi^+$). Then the two flavours of neutrinos emerge from UHECR hadronic interactions, as the pion decays further with emission of a muon-neutrino ($\pi^+ \rightarrow \mu^+ + \nu_\mu$) in the primary decay mode, and the muon decays further creating an electron-neutrino ($\mu^+ \rightarrow e^+ + \overline{\nu_\mu} + \nu_e$), as discussed for example in \citet{Kun2013}. The muon-neutrinos imprint track-type events, while the electrons imprint shower-like events on the detector array of the IceCube. For the pion-production, the proton speed has to be relativistic with high Lorentz-factor, which is common inside a newly formed jet as presented above. Then the high Lorentz factor is the most likely condition to explain a correlation between HE neutrino events and compact flat-spectrum radio AGN with a flat spectrum extending to very high frequency.

\section{Discussion and summary}
\label{section5}

We cross-correlated the Parkes Catalogue and the Second Planck Catalogue of Compact Sources in order to find possible candidates for the $15$ track-type neutrino events detected by the IceCube detector. We found four flat-spectrum radio sources close enough to a track-type neutrino event, two of them being also detected at high frequencies by \textit{Planck}.

Next we estimate the treble chance-coincidence of finding a flat spectrum radio source (from the Parkes Catalogue), at higher frequencies (from the Planck Catalogue) within the error-box of the track events on the sky. For this we employ the results of \citet{Drinkwater1997}, who by using the Parkes Catalogue identified $323$ flat-spectrum sources ($\alpha_\mathrm{2.7GHz,5GHz}>-0.5$) in an area of $3.9$ sr of the sky. Assuming a homogeneous distribution, this translates to $\sim1040$ flat-spectrum radio sources over the full sky ($=41252$ deg$^2$). Taking $1.{\degr}2$ as the average median angular error of the track-type neutrino events, and then the average area of $4.52$ deg$^2$ of one flat spectrum radio source, the statistics yields $\sim0.11$ such source over neutrino event area. A fraction of about 1/10 of flat-spectrum sources defined near 5 GHz extend to high Planck frequencies \citep{PCCS2015}, which means an $\sim0.01$ flat-spectrum source over neutrino event area. The combined probability for two flat-spectrum sources to emerge as candidate for track-type HE neutrino events by chance has then the tiny value of $10^{-4}$. We may conclude that the coincidence is very probably real. As for other track-type HE neutrino events, it is plausible that they may pertain to sources at yet higher red-shift, and so at radio flux densities is below the present detection threshold.

We presented the flat-spectrum blazar PKS~0723--008 as an excellent candidate for ID5. We analysed the available MOJAVE data of the sources, which led to the selection of PKS~0723--008 from the two sources appearing at \textit{Planck} frequencies. This is a source with three important characteristics: it has flat radio spectrum at high frequencies, its radio flux significantly increased in the last decade, and it is within the median angular error of a track-type HE neutrino event (ID5). The spectrum brightened and flattened in the \textit{Planck} frequencies after 2006, with a local maximum in the integrated flux density by 2011, suggesting a violent process in the core. The radio maps after 2011 reveal a component ejection from the core (Fig. \ref{figure:vlba}), explaining the local maxima in the total flux density. Such flares in the radio wavelengths are attributed to adiabatic shock-in-jet models \citep{Marscher1985,Hughes1985}. Alternatively, the Turbulent Extreme Multi-Zone Model of \citet{Marscher2014} also increases the flux density of the source, when the magnetically turbulent ambient jet flow crosses oblique or cone-shaped shocks. However, such a scenario seems disfavoured in the present case due to the observed component ejection. We argued that the probability for an accidental identification of two flat-spectrum sources extending their spectrum to near THz frequencies with track-type HE neutrino events is extremely small ($\sim 10^{-4}$).

 We proposed a scenario explaining the track-type HE neutrino event through a binary SMBH, which upon merging induces the reorientation of the new jet towards Earth, providing a strong boosting of all emissions. The scenario predicts low-frequency gravitational waves, UHECR, HE neutrinos, and luminous radio afterglow with flat spectrum extending to near THz frequencies, all generated by the merger of two SMBHs acting as engine.

\section*{Acknowledgements}

The authors thank the referee for the suggestions helping to improve this Letter. EK acknowledges the discussions with Krisztina \'{E}. Gab\'{a}nyi (F\"{O}MI Satellite Geodetic Observatory,). PLB acknowledges the discussions on this topic with Ben Harms (UA Tuscaloosa), Francis Halzen (U. Wisconsin), Karl Mannheim (U. W{\"u}rzburg), Athina Meli (U. Gent), Julia Becker Tjus (U. Bochum), and Dawn Williams (U. Alabama). This research has made use of data from the MOJAVE data base maintained by the MOJAVE team \citep{Lister2009} and of the NASA/IPAC Extragalactic Database (NED), operated by the Jet Propulsion Laboratory, California Institute of Technology, under contract with the National Aeronautics and Space Administration.

\end{document}